\documentclass[journal]{IEEEtran}
\usepackage{amsmath,amsfonts}
\usepackage{algorithmic}
\usepackage[ruled,vlined]{algorithm2e}
\usepackage{array}
\usepackage[colorlinks=true, linkcolor=red, citecolor=green]{hyperref}
\usepackage{textcomp}
\usepackage{stfloats}
\usepackage{url}
\usepackage{verbatim}
\usepackage{graphicx}
\usepackage{cite}
\usepackage{upgreek}
\usepackage{xcolor}
\usepackage{amssymb}
\usepackage{subfigure}
\usepackage{makecell}
\usepackage[capitalize,noabbrev]{cleveref}
\usepackage{algorithmic}
\begin{document}

\title{Hierarchical Multi-Agent Reinforcement Learning-based Coordinated Spatial Reuse for Next Generation WLANs}

\author{Jiaming Yu,~\IEEEmembership{Graduate Student Member,~IEEE,} Le Liang,~\IEEEmembership{Member,~IEEE,} Hao Ye,~\IEEEmembership{Member,~IEEE,}\\ and Shi Jin,~\IEEEmembership{Fellow,~IEEE}
\thanks{Jiaming Yu, Le Liang, and Shi Jin are with the National Mobile Communications Research Laboratory, Southeast University, Nanjing 210096, China
(e-mail: jiaming@seu.edu.cn; lliang@seu.edu.cn; jinshi@seu.edu.cn). Le Liang is also with Purple Mountain Laboratories, Nanjing 211111, China.}
\thanks{Hao Ye is with the Department of Electrical and Computer Engineering, University of California, Santa Cruz, CA 95064, USA (e-mail: yehao@ucsc.edu).}
}

\maketitle

\begin{abstract}
High-density Wi-Fi deployments often result in significant co-channel interference, which degrades overall network performance. To address this issue, coordination of multi access points (APs) has been considered to enable coordinated spatial reuse (CSR) in next generation wireless local area networks. This paper tackles the challenge of downlink spatial reuse in Wi-Fi networks, specifically in scenarios involving overlapping basic service sets, by employing hierarchical multi-agent reinforcement learning (HMARL). We decompose the CSR process into two phases, i.e., a polling phase and a decision phase, and introduce the HMARL algorithm to enable efficient CSR.
To enhance training efficiency, the proposed HMARL algorithm employs a hierarchical structure, where station selection and power control are determined by a high- and low-level policy network, respectively. Simulation results demonstrate that this approach consistently outperforms baseline methods in terms of throughput and latency across various network topologies.
Moreover, the algorithm exhibits robust performance when coexisting with legacy APs.
Additional experiments in a representative topology further reveal that the carefully designed reward function not only maximizes the overall network throughput, but also improves fairness in transmission opportunities for APs in high-interference regions.
\end{abstract}


\section{Introduction}
Wi-Fi has become a pivotal technology in wireless local area networks (WLANs), with the latest commercial technologies Wi-Fi 6 \cite{khorov2018tutorial} and Wi-Fi 7 \cite{deng2020ieee} widely deployed in various scenarios to provide users with high data rate coverage. Looking forward, the next generation Wi-Fi standard, i.e., Wi-Fi 8 \cite{reshef2022future, galati2024will}, aims to further improve the experience of intelligent connectivity in ultra-dense network environments.
In such environments, a key factor that influences the network performance is the medium access control mechanism.
As the most popular access mechanism, the carrier-sense multiple access with collision avoidance (CSMA/CA) mechanism plays a critical role in managing how devices share the wireless medium, which abides by the listen-before-talk protocol. 
{Specifically, a user initiates transmission only when it senses that the received signal power is below a predefined clear channel assessment (CCA) threshold, indicating that the channel is sensed idle.}
However, in densely deployed access point (AP) environments, CSMA/CA leads to severely reduced transmission opportunities (TXOP) for nodes in the overlapping basic service set (OBSS) \cite{shin2015per, iwata2019analysis}. 
This degradation occurs because the nodes frequently detect the channel as busy due to co-channel interference, preventing them from initiating transmissions. 
\begin{figure}[!t]
\centering
\includegraphics[width=0.45\textwidth]{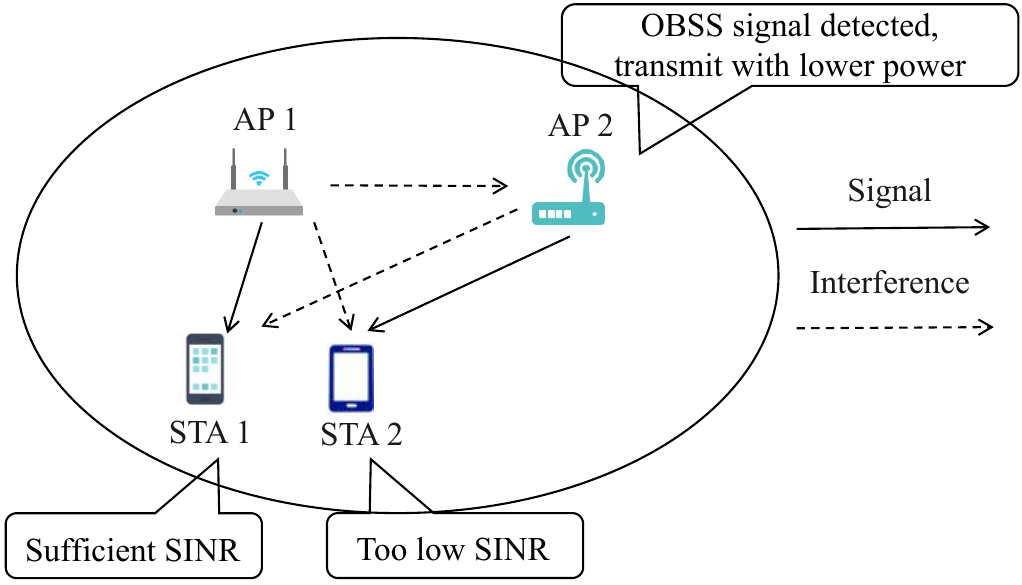}
\caption{An example of BSS coloring. AP 1 first detects that the channel is idle and transmits with maximum power. Subsequently, AP 2 detects the signal from AP 1 and transmits simultaneously but at a lower power. This leads to STA 2 receiving an excessively low SINR, causing the transmission to fail.}
\label{obss coloring}
\end{figure}

To improve the performance in high-density environments, several studies analyze the performance of OBSS scenarios and propose specific mechanism to improve the network throughput.
For example, the CSMA/CA mechanism has been evaluated to suffer severe performance degradation in densely deployed environments \cite{abinader2014performance}. In response, an AP association optimization algorithm has been proposed in \cite{oni2015ap} that matches stations (STAs) with APs based on the signal-to-interference-plus-noise ratio (SINR). This approach adjusts the CCA threshold to increase concurrent transmissions, i.e., spatial reuse \cite{wilhelmi2021spatial}.
{Furthermore, the basic service set (BSS) coloring technology \cite{wifi6standard} has been introduced in Wi-Fi 6 to adjust a dynamic CCA threshold.
Specifically, each AP attaches a color identifier to its transmitted packets. When an AP receives a packet with a different color, indicating that it belongs to a neighboring BSS, it applies a higher CCA threshold to determine whether the channel is sensed idle. The high CCA threshold increases the opportunity for spatial reuse. At the same time, the AP also needs to reduce the transmit power to minimize interference to co-channel STAs. 
However, as illustrated in Fig~\ref{obss coloring}, reducing the transmit power may lead to transmission failures because the AP focuses only on mitigating interference to other STAs while neglecting the SINR of its own associated STA, resulting in potential transmission failure.
This is because BSS coloring neglects the spatial interference patterns arising from the relative positions of APs and STAs, limiting the overall network performance.}



To solve this problem, several studies consider optimizing the transmit power to facilitate spatial reuse. 
In \cite{nunez2022txop}, a transmission scheduling mechanism for APs has been proposed to achieve spatial reuse. 
However, a controller for APs needs to obtain the received signal strength indicators of each STA and then utilizes a brute-force approach to traverse all combinations of discrete power control to obtain the highest throughput, thus increasing the complexity of the algorithm.
Moreover, a centralized AP scheduling scheme has been proposed in \cite{tuan2023improving} to reduce the transmit power of the interference and extend the TXOP with an AP controller that maintains the overlapping area record of each OBSS. 
Furthermore, a novel Bayesian optimization algorithm based on Gaussian process has been further developed in \cite{liu2024bayesian} to jointly optimize the CCA threshold and transmit power.

In addition to traditional methods, recent research has explored artificial intelligence-based approaches to improve the network performance in OBSS scenarios \cite{iturria2023meta, iturria2024cooperate}, with a particular focus on deep reinforcement learning (RL).
In \cite{zhang2022multiaccess}, an RL-based channel access protocol has been proposed, where a centralized AP controller assigns channels to each AP as an agent, and then each AP performs channel access. 
Moreover, a similar problem has been solved by extracting topology features via a graph convolution network as observations of the RL agent for channel allocation \cite{nakashima2020deep}.
However, these works are mainly aimed at the multi-channel allocation problem for APs in OBSS scenarios, while other works consider applying RL algorithms for spatial reuse on a single frequency band. 
In \cite{zhang2023ddpg}, the information of other APs, such as the packet loss rate, is collected by a master AP to optimize their contention window values and CCA thresholds of each AP using the deep deterministic policy gradient algorithm \cite{lillicrap2015continuous}, thereby improving the total throughput in the OBSS scenario.
A similar joint optimization of the contention window values and CCA thresholds problem has been considered in \cite{yan2024multi} to reduce the network latency, where each AP broadcasts its own information at regular intervals and receives information from other APs for decision making without any centralized controllers. 
Moreover, a deep Q network \cite{mnih2015human} has been utilized to jointly optimize the selection of AP CCA threshold and transmit power control to improve the capacity \cite{huang2022deep}. This structure has been further extended in \cite{huang2023hierarchical}, where each AP uses a hierarchical RL algorithm at two different levels to optimize the CCA threshold and the transmit power, respectively, minimizing the length of packet queues in the network. 

{Although the aforementioned studies have contributed to improving spatial reuse, a fundamental challenge remains. Specifically, most existing approaches rely on dynamically adjusting the CCA threshold to enable spatial reuse, which is a passive method, i.e., the AP transmits only when it detects that the received power is below the CCA threshold.
As a result, in dense AP deployments, APs located in highly interfered areas may detect the channel as busy most of the time, even with an elevated CCA threshold.
This prevents them from actively exploiting spatial reuse opportunities, leading to reduced throughput and increased latency, which do not satisfy the requirements of low-latency applications such as virtual reality/augmented reality, remote surgery, and online gaming.
To overcome this problem, multi-AP cooperation technologies \cite{verma2023survey} have been considered as a key feature in next generation WLANs, which enable proactive spatial reuse by coordinating downlink transmissions and optimizing transmit power \cite{zhu2025two, wang2024research}, thereby promoting coordinated spatial reuse (CSR).
However, existing research for CSR has some limitations. 
On the one hand, optimizing only the transmit power is insufficient \cite{zhu2025two, wang2024research}, as slight mismanagement of transmit power in dense deployment scenarios can lead to transmission failures that result in significant fluctuations in overall network performance.
Therefore, the joint optimization of STA selection and power control is crucial because the success of CSR depends heavily on the spatial topology of the STAs.
On the other hand, they assume that a centralized controller has a wired connection to the APs, which takes data from APs and performs CSR configuration \cite{wojnar2024ieee, wojnar2025coordinated}. The centralized approach requires the APs to be located in a single management domain, such as a factory and university campus. However, centralized controllers are not always available, especially in residential scenarios, thus limiting the practical applicability of such approaches.
Motivated by the above limitations, we propose a hierarchical multi-agent reinforcement learning (HMARL) algorithm, to solve the downlink spatial reuse problem in this paper. The proposed HMARL algorithm is fully decentralized and does not require a central AP controller, enabling coordination among APs in diverse deployment scenarios.} Specifically, we decompose the CSR process into two phases, a polling phase and a decision phase. The polling phase enables APs to proactively perform spatial reuse, while also facilitating the share of information among APs for better decision making. 
As for the decision phase, we design a hierarchical multi-agent reinforcement learning (MARL) architecture where each AP maintains a high-level policy to select which STA to transmit to in the downlink transmission and a low-level policy to perform the power control to enable the CSR. In addition to maximizing overall throughput, our design also considers fairness among APs, even if improving fairness slightly sacrifices the overall performance of the network.
The contributions and main results of this work are summarized as follows.

\begin{itemize}
\item To effectively address the CSR problem, we divide the CSR process into two phases, the polling phase and the decision phase, and design a corresponding framework to coordinate APs for the CSR.

\item We propose a hierarchical MARL algorithm for the CSR problem in next-generation WLANs. Considering that the centralized training by a centralized node is impractical in real OBSS scenarios, especially in residential scenarios, HMARL adopts a fully distributed paradigm for training and execution. 

\item  We provide extensive simulation results to demonstrate the improvements of our framework in network throughput and fairness. We also conduct ablation studies to assess the effectiveness of broadcast information and hierarchical architecture, as well as robustness evaluations under varying network conditions.
\end{itemize}

The remainder of this paper is structured as follows. The system model and problem formulation are described in Section II.
Section III provides the hierarchical MARL preliminaries and the underlying decentralized partially observable Markov decision process (Dec-POMDP) model.
We present the proposed HMARL algorithm and protocol in Section IV. Finally, simulation results are discussed in Section V, followed by the conclusion and discussion of the future work in Section VI.

\section{System Model and Problem Formulation}
\begin{figure}[!t]
\centering
\includegraphics[width=0.45\textwidth]{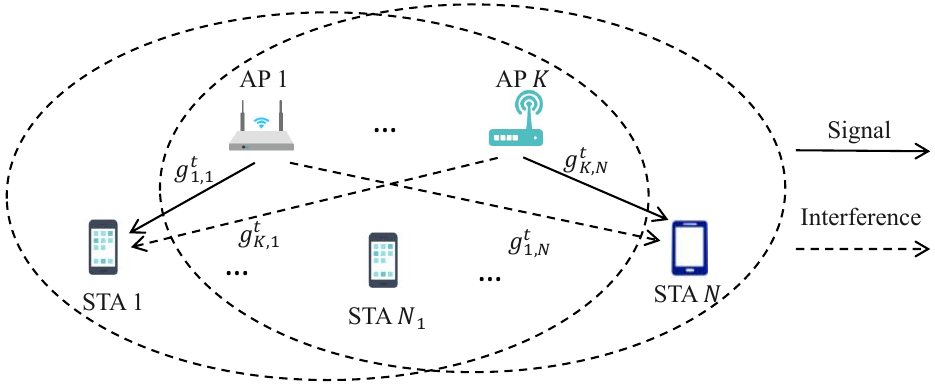}
\caption{Scenario of a downlink OBSS.}
\label{obss}
\end{figure}
\begin{figure*}[!t]
\centering
\includegraphics[width=0.95\textwidth]{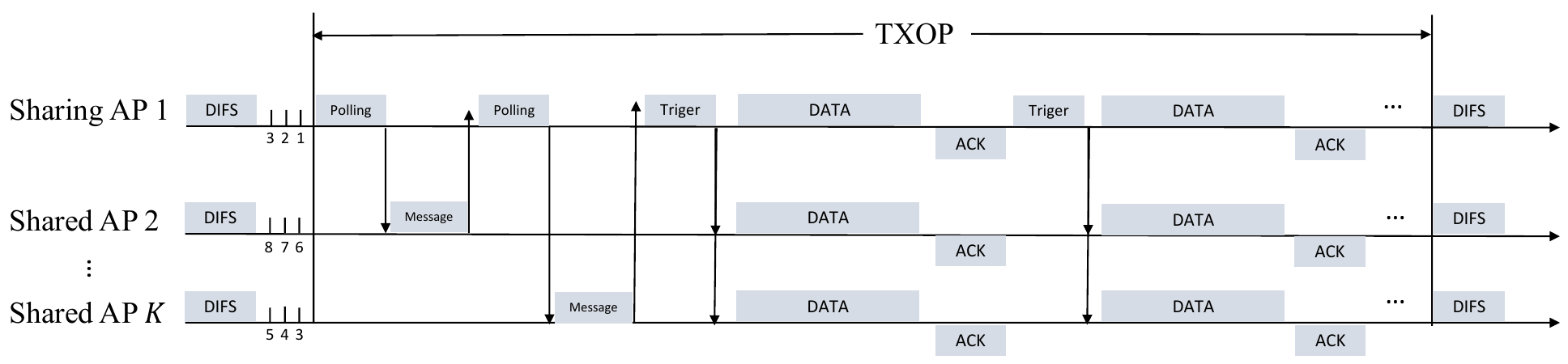}
\caption{Operation of the CSR scheme.  APs still utilize the CSMA/CA protocol for channel competition. The AP that occupy the channel becomes the sharing AP and share the channel with other APs for spatial reuse.}
\label{CSR}
\end{figure*}
\subsection{CSR System Model}

As illustrated in Fig.~\ref{obss}, we consider the downlink distributed channel access problem in a time-slotted OBSS scenario consisting of $K$ BSS, each containing $N_1,N_2,...,N_K$ STAs, respectively. AP $k$ is equipped with $N_k$ finite queue buffers and attempts to communicate with its respective STAs. Packet arrivals follow a predefined distribution, and packets are dropped when the buffer is full. To remain compatible with legacy Wi-Fi protocols, each AP retains the CSMA/CA mechanism to contend for a TXOP to transmit. However, in dense deployments, frequent channel contention and co-channel interference drastically reduce the likelihood of successful TXOP acquisition, particularly under saturated traffic conditions, thereby degrading system performance.

To overcome the inherent limitation of the traditional CSMA/CA mechanism, our work focuses on the CSR mechanism aimed at improving simultaneous transmissions. As illustrated in Fig.~\ref{CSR}, APs continue to contend for TXOPs using CSMA/CA. 
Specifically, APs first detect the channel as idle and then wait for a distributed inter frame space (DIFS) interval before contending for the channel through the binary exponential backoff algorithm.
If only one AP successfully acquires the TXOP, it becomes the sharing AP. Conversely, if two or more APs attempt to occupy the TXOP, a collision occurs and no AP becomes the sharing AP. Then they continue adopting the CSMA/CA mechanism to contend for the TXOP until only one AP successfully contends and becomes the sharing AP. Upon acquiring the TXOP, the sharing AP initiates the CSR process by polling other APs—referred to as shared APs—to determine their willingness to participate in CSR during the current TXOP duration. Subsequently, the APs that decide to participate in the CSR process send their statistical information to the sharing AP, which is introduced in detail in Section IV. For simplicity, we assume that all APs with non-empty buffers participate in the CSR process. Therefore, the relationship between APs in the OBSS are both competitive and cooperative.
Once the polling completes, the sharing AP broadcasts the collected information to each AP via a trigger frame, instructing shared APs to take an action as an RL agent to start transmission. 
{Specifically, since the success of the CSR heavily depends on the spatial topology of STAs, careful STA selection is critical for minimizing interference. Therefore, each shared AP selects which STAs to transmit to. Meanwhile, to avoid consistently excluding STAs in strong interference positions, the sharing AP adopts a round-robin method to send packets to its respective STAs.} Due to the role of the sharing AP being determined by the CSMA/CA protocol, each AP has an opportunity of becoming the sharing AP and sending packets to all associated STAs. 
Furthermore, to further reduce interference, each AP adjusts its transmit power through power control. 
Therefore, this approach minimizes interference and ensures that each STA can receive packets.
After an AP transmits, the transmission is not successful until the corresponding acknowledgment (ACK) frame is received. To reduce the communication overhead between APs, each AP can transmit multiple packets in a TXOP duration. 

\subsection{Problem Formulation}
We assume that the channel fading remains constant during a packet transmission. During coherence time $t$, the channel power gain, $g_{k,i}^t$, from AP $k$ to STA $i$ follows
\begin{equation}
g_{k,i}^t=\alpha_{k,i}^th_{k,i}^t,
\end{equation}
where $h_{k,i}^t$ represents the small-scale fading power component and $\alpha_{k,i}^t$ captures the large-scale fading effect, including path loss and shadowing.
The received SINR $\gamma_{k,i}^t$ when AP $k$ transmits to STA $i$ at time slot $t$ is expressed as
\begin{equation}
\gamma_{k,i}^t=\frac{p_k^tg_{k,i}^t}{\sum\limits_{k'\neq k}p_{k'}^tg_{k',i}^t+\sigma^2},
\end{equation}
where $p_k^t$ is the transmit power of AP $k$ at time slot $t$, and $\sigma^2$ is the noise power. 
{The transmission success probability is determined by both the modulation and coding scheme (MCS) and the SINR. Specifically, a transmission is considered successful if the received SINR exceeds the threshold associated with the selected MCS, enabling successful packet decoding at the receiver \cite{krotov2020rate}.} In this work, for simplicity, we assume a fixed MCS across all APs. Accordingly, we define a binary indicator function $I(\gamma_{k,i}^t)$, where $I(\gamma^t_{k, i})=1$ denotes a successful transmission from AP $k$ to STA $i$ at time $t$, and 0 otherwise.
The objective of the CSR problem is to maximize network throughput while maintaining fairness among each AP with performing transmission STA selection and power control in a TXOP duration. The problem can be formulated as
\begin{equation}
\begin{aligned}
& \mathop{\max}_{\mathbf{P}_t, \mathbf{x}_t} &&\sum_{t=0}^T\sum_{k=1}^K\sum_{i=1}^{N_k} \frac{x_{k,i}^tI(\gamma_{k,i}^t)}{T}, \\
&\ \text{s.t.} && \beta_k^{t_0}\geq \beta_\text{min}, \quad\forall k,t_0,\\ 
&&&0\leq P_k^t\leq P_{\max},\quad\forall k,t,\\
&&& x_{k,i}^t \in \{0,1\},\quad\forall k,i,t.\\
&&&\sum_{i=1}^{N_k}x^t_{k,i}\leq1, \forall k,
\label{problem}
\end{aligned}
\end{equation}
where $\beta_k^{t_0}=\sum_{t=t_0}^{t_0+T_0}\sum_{i=1}^{N_k}\frac{x_{k,i}^tI(\gamma_{k,i}^t)}{T_0}$ represents the instantaneous throughput for AP $k$ over a sliding window of length $T_0$, starting from time $t_0$. The first constraint indicates that each AP, even in the position with the strongest interference, should achieve a certain level of throughput $\beta_\text{min}$ to ensure fairness across APs. Furthermore, $\mathbf{P}_t=[P_1^t, P_2^t, ... , P_K^t]$ and $\mathbf{x}_t=~ [x_{1,1}^t,...,x_{1,N_1}^t,x_{2,1,}^t,...,x_{2,N_2}^t,...,x_{K,N_K}^t]$ denote the transmit power vector for APs and the vector for STAs selection at time $t$, respectively. Note the AP can only transmit to at most one STA at a time, i.e., $\sum_{i=1}^{N_k}x^t_{k,i}\leq1, \forall k$. Moreover, $P_\text{max}$ is the maximum transmit power for all links between AP and STA.

\section{Hierarchical MARL Preliminaries and Dec-POMDP Model}
\label{Section III}
The aforementioned CSR problem can be formulated as a MARL problem where each AP makes decisions based solely on its own observations and received information. Due to the dynamic nature of AP deployments, there is no centralized node for training. Consequently, the training and deployment of the MARL problem are distributed, which can be modeled as a Dec-POMDP model. Mathematically, the underlying Dec-POMDP model can be described by a tuple $<\mathcal{K}, \mathcal{M}, \{\mathcal{A}\}^i_{i\in\mathcal{K}}, \{\mathcal{O}\}^i_{i\in\mathcal{K}}, \mathcal{Z},\mathcal{P},R,\gamma>$, where $\mathcal{K}$ denotes the set of agents (i.e., APs), with $|\mathcal{K}| = K$, and $\mathcal{Z}$ represents the observation space. The feature space $\mathcal{M}$ is extracted from the observations in $\mathcal{Z}$. Each agent $i$ selects actions from the action space $\mathcal{A}_i$ and options from the option space $\mathcal{O}_i$, with the environment dynamics governed by the transition probability $\mathcal{P}$ and reward function $\mathcal{R}$. The discount factor $\gamma \in (0,1]$ controls the weighting of future rewards. 

At each time step $t$, agent $i$ obtains its local observation $z^i_t\in \mathcal{Z}$ from the current environment. If the time step $t$ is in the polling stage, agent $i$ extracts the feature $m^i_t\in\mathcal{M}$ from the observation $z^i_t$ and transmits it to the sharing AP, and then the sharing AP aggregates the messages and broadcasts them to the shared APs. Upon receiving the broadcast message, each shared AP selects an option $o^i_t\in\mathcal{O}^i$ using a high-level policy. Subsequently, whenever an agent attempts to send a packet, it applies a low-level policy to select an action $a^i_t\in\mathcal{A}^i$ based on its local observation, the shared message and the selected option, resulting a joint action $\boldsymbol{a}_t\in\mathcal{A}$, which is introduced in detail in Section IV. The environment then returns to each agent a reward $r_t\in R$ and transitions to the next state under the state transition probability 
$\mathcal{P}$
of the MDP. The objective of each agent is to learn both the its high-level and low-level policies to maximize the expected discounted return $\mathbb{E}[\sum_{t=0}^\infty\gamma^tr_t],$ where $\gamma \in (0, 1]$ is a discount factor. The corresponding RL elements for the CSR problem are introduced hereafter.

\textbf{Option and Action:} At the beginning of each TXOP, each shared AP selects an option $o^i_t \in \{1,2,...,N_i\}$ based on its high-level policy after receiving the information broadcast by the sharing AP. The option indicates to which STA the AP will transmit during the entire TXOP. Note that the sharing AP takes the option determined by its buffer at each time step $t$. Furthermore, each AP can take an action $a^i_t\in\{P_1,P_2,...,P_\text{max}\}$ as the power control at each packet transmission to mitigate the mutual interference, where $P_\text{max}$ denotes the maximum power.

\textbf{Local Observation:} The observation of agent $i$ at time step $t$ is defined as $z^i_t\triangleq\{a_{t-1}^{i}, \boldsymbol{l}_{t}, d_{t},s_{t}^{i},\boldsymbol{q}_{TX-1}^{i}, \boldsymbol{c}_{TX-1}^{i}\}$, where $a^i_{t-1} $ represents the action taken at the previous time step. Since the key to a successful spatial reuse is the topological position of the STAs, we assume that each AP knows the relative positions of its respective STAs, denoted by $\boldsymbol{l}_t$, which can be determined by Wi-Fi sensing technologies \cite{zhang2022wi,ratnam2024optimal,du2024overview}. Additionally, $d_t$ denotes the index of the STA to which the sharing AP attempts to transmit at time step $t$, which can be broadcast in the trigger frame to shared APs. Broadcasting $d_t$ is crucial, as the sharing AP can not disclose the topological position of the STA due to privacy concerns. Thus, the sharing AP can only broadcast the index number of its respective STAs to implicitly announce the STA topological location feature. Furthermore, $s^i_{t-1}$ represents the SINR of AP $i$ at time step $t-1$, which can be detected by receiving the previous ACK frame. Moreover, the remaining terms $\boldsymbol{q}_{TX-1}^{i}$ and $\boldsymbol{c}_{TX-1}^{i}$ refer to the statistic vectors in the previous TXOP. Specifically, $\boldsymbol{q}_{TX-1}^{i}=[q_1, q_2,...,q_{N_i}]$ records the number of packets that AP $i$ successfully transmits to its respective STAs in the last TXOP, where $N_i$ and $TX$ are the number of the respective STAs and the index of the TXOP at time step $t$, respectively. Similarly, $\boldsymbol{c}_{TX-1}^{i}=[c_1^1, c_1^2,...,c_1^{E_0}, c_2^1,c_2^2,..., c_{N_i}^{E_0}]$ records the transmission situation of AP $i$ in the previous TXOP, where $E_0$ is the number of transmissions per TXOP. For example, $c_{N_i}^{E_0}=1$ if AP $i$ successfully transmits a packet to STA $N_i$ at the $E_0$th transmission within the TXOP, $c_{N_i}^{E_0}=2$ if the transmission fails, and $c_{N_i}^{E_0}=0$ if AP $i$ does not transmit to the STA. To enhance decision-making, the historical local observation for AP $i$ at time step $t$ is represented as $\tau^i_t\triangleq[z^i_{t-M+1},...,z^i_{t-1}, z^i_{t}]$, where $M$ is the length of the local observation history. 

\textbf{Reward:} Reward design is critical for improving the overall system performance. In the investigated CSR problem described in Section II, our objectives are twofold: maximizing the network throughput while ensuring as much fairness as possible among each BSS.

In response to the first objective, the reward could be defined as the total number of the successfully transmitted packets, $\sum_{k=1}^K\sum_{i=1}^{N_k} x_{k,i}^tI(\gamma_{k,i}^t)$, at each time step $t$. However, this reward design is against the second objective due to a lack of consideration of fairness. To address this, we define a total reward function $r_{t,\text{tot}}$ that balances throughput and fairness.
Specifically, the reward is given by
\begin{equation}
\label{r_tot}
r_{t,\text{tot}}=\sum\limits_{i\in\mathcal{S}_t}log(\frac{1}{u^i_t}),
\end{equation}
where $\mathcal{S}_t$ denotes the set of APs that successfully transmit at time step $t$, and $u^i_t$ represents throughput from the previous period of time of agent $i$. The total reward promotes the behavior of successful transmission by assigning positive rewards. 
In particular, the negative logarithmic relationship between the reward and throughput implies that APs with lower past throughput receive higher rewards for successful transmission. As a result, the reward mechanism inherently prioritizes underperforming agents, improving their transmission opportunities.
However, relying solely on the total reward may still lead to suboptimal solutions. For example, APs with good topological locations may always choose to transmit alternately at maximum power, whereas APs with poor topological locations may never transmit. In this case, each AP can still obtain a satisfactory reward due to the positive reward value computed by (\ref{r_tot}). As a result, agents have little incentive to explore better joint actions that could potentially improve overall network performance. Therefore, independent rewards need to be designed for each agent to ensure fairness finely. Specifically, the individual reward $r^i_{t,\text{ind}}$ for agent $i$ at time step $t$ is 
\begin{equation}
r_{t,\text{ind}}^i=
\begin{cases}
-\frac{a_t^i}{P_{\max}}(|\mathcal{C}_t|-1), & \text{if}~i\in\mathcal{C}_t, \\
-\frac{a_t^i}{P_{\max}}|\mathcal{C}_t|, & \text{otherwise}, \\
\end{cases}
\end{equation}
where $\mathcal{C}_t$ and $a^i_t$ represent the set of APs that collide at time step $t$, and action of agent $i$, i.e., the transmit power, respectively. The individual rewards means that, when a collision occurs, each agent receives a negative reward that is proportional to its transmit power. The basic idea of the individual rewards is to more heavily penalizes nodes with higher transmit power because they are more likely to cause substantial interference, making it difficult for other STAs to successfully receive packets. By encouraging lower transmit powers in the event of collisions, the individual rewards aim to reduce interference and promote more equitable transmission across all nodes, ultimately leading to improved network efficiency and performance.

Building upon the definitions of the total and individual rewards, the reward at each time step t is formally constituted as a vector: $r_t~\triangleq~[r_{t,\text{tot}}, r^1_{t,\text{ind}},...,r^K_{t,\text{ind}}]$.

\section{HMARL Algorithm and Protocol}
Based on the DEC-POMDP formulation introduced above, we propose a fully distributed hierarchical MARL algorithm, where multiple APs collectively explore the environment and improve the transmission selection of STAs and power control strategies based on their own observations of the environment state and the information shared by others. We begin by presenting the two phases of a TXOP, i.e., the polling phase and the decision phase, followed by the proposed HMARL algorithm utilized to train each agent.
\subsection{Polling Phase}
When an AP needs to transmit a packet, it competes for the TXOP to become the sharing AP through the CSMA/CA mechanism. Once successful, it initiates the polling phase to coordinate with other APs. During the polling phase, AP $i$ participating in CSR sends its information of extracted by local observations to the sharing AP, who then shares the collective information with all APs, enabling more informed and coordinated decision-making. This approach mitigates the issue of environmental non-stationarity that can arise when agents rely solely on local observations and that may lead to suboptimal decisions. Without such information sharing, each AP would independently optimize its actions based on limited observation, potentially increasing interference and reducing overall network performance. 

\begin{figure}[!t]
\centering
\includegraphics[width=0.49\textwidth]{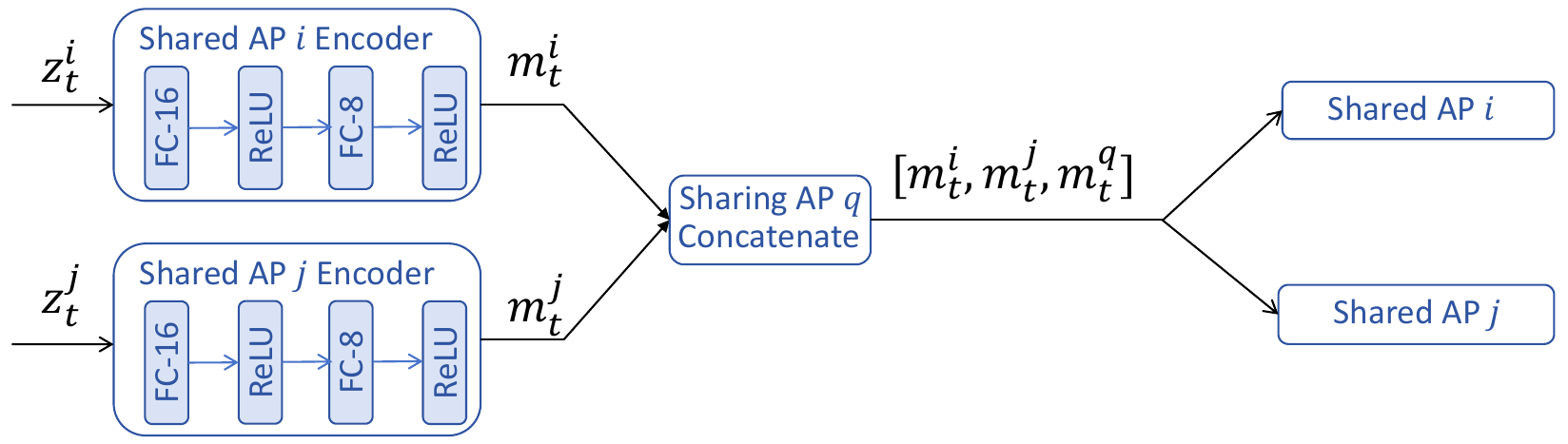}
\caption{The message encoder structure: Each participating AP uses an encoder to compress its observation into a compact message, and then forwards this message to the sharing AP. The sharing AP collects these compressed messages from shared APs, concatenates them with its own information, and broadcasts the combined data to all APs.}
\label{polling phase}
\end{figure}

Nevertheless, since the observation of each AP contains information about the relative locations of the associated STAs, direct sharing of raw observations raises privacy concerns and incurs significant communication overhead. To address these issues, each AP employs a neural encoder to compress its observation before transmission. As illustrated in Fig.~\ref{polling phase}, the encoder of each AP consists of two fully connected hidden layers, containing 16 and 8 neurons, respectively. The rectified linear unit (ReLU), $f(x)=\text{max}(0,x)$, is utilized after each layer of neurons as the activation function. Each shared AP utilizes the encoder to compress the observation $z^i_t$ into information $m^i_t$, i.e., 
\begin{equation}
m^i_t=e(z^i_t;\boldsymbol{\theta}^i_\text{m}),
\end{equation}
where $e(\cdot)$ denotes the encoder network with parameters $\boldsymbol{\theta}^i_\text{m}$.
These encoded messages are sent to the sharing AP, which concatenates them with its own and broadcasts the aggregated message to all APs.

\subsection{Decision Phase}
\begin{figure*}[!t]
\centering
\includegraphics[width=0.95\textwidth]{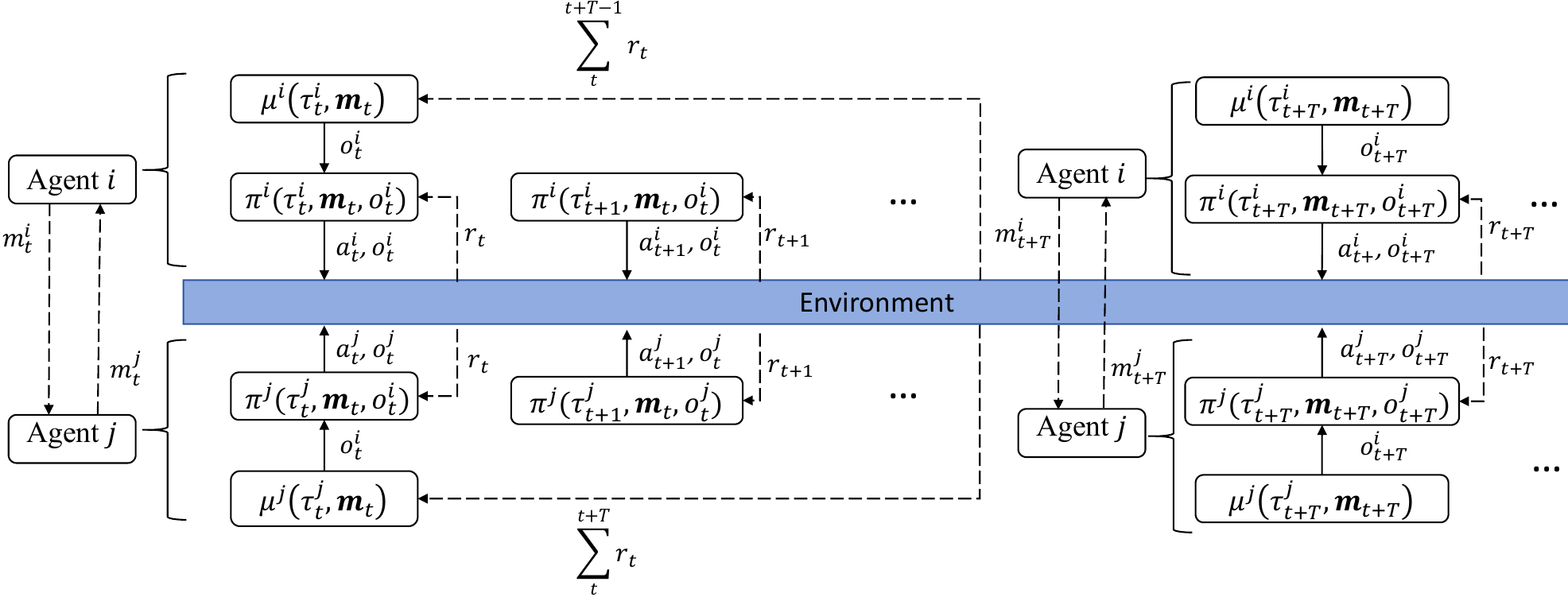}
\caption{Decision phase of a CSR process. The high-level policy selects which STA to transmit and the low-level policy selects the transmit power.}
\label{decision phase}
\end{figure*}
At the end of the polling phase, the sharing AP broadcasts a trigger frame to the shared APs to initiate the decision phase. In this phase, each AP utilizes the received information and its local observation to make decisions on transmission STA selection and power control. To mitigate the exploration complexity arising from a large action space, we adopt a hierarchical decision-making structure. As illustrated in Fig.~\ref{decision phase}, AP $i$ has two policy networks, i.e., the high-level policy network $\mu_i$ and the low-level policy network $\pi_i$, which are used to make decisions on transmission STA selection and power control, respectively. Specifically, multiple packet transmissions are allowed within a single TXOP duration. At the beginning of the decision-making phase, the high-level policy utilizes both its local observation history $\tau^i_t$ and received information history $\boldsymbol{m}_t=\text{Concat}([m_{t-M}^i,m_{t-M+1}^i...,m_{t-1}^i,m_t^i],\forall i\in\mathcal{K})$ as inputs to determine the transmission STA selection $o^i_t$, which can be represent as 
\begin{equation}
o^i_t=\mu_i(\tau^i_t,\boldsymbol{m}_t;\boldsymbol{\theta}^i_\mu),
\end{equation}
where $\boldsymbol{\theta}^i_\mu$ denotes the parameters of agent $i$ high-policy $\mu_i$. Once the STA is selected, it remains fixed during the entire TXOP duration. Subsequently, the low-level policy $\pi_i$ takes power control over each packet transmission, i.e.,
\begin{equation}
a^i_t=\pi_i(\tau^i_t,\boldsymbol{m}_t,o^i_t;\boldsymbol{\theta}^i_\pi),
\end{equation}
where $\boldsymbol{\theta}^i_\pi$ represents the parameters of agent $i$ low-policy $\pi_i$.

\subsection{HMARL Algorithm}
Based on the distributed decision framework introduced above, we propose to adopt the proximal policy optimization (PPO) algorithm \cite{schulman2017proximal}—a widely recognized policy-based reinforcement learning algorithm—to train the high- and low-level policy networks for each agent.
In policy-based algorithms, agent learns a policy $\pi$ that maximizes the expected discounted return 
\begin{equation}
\mathbb{E}_\pi[\sum_{t=0}^T\gamma^tr_t]=\sum_\tau R(\tau)P(\tau),
\end{equation}
where $\tau=\{s_{0},a_{0},r_{0},s_{1},a_{1},r_{1},...,s_{T},a_{T},r_{T}\}$ represents a trajectory under its current policy $\pi$. Moreover, $R(\tau)=\sum_{t=0}^T\gamma^tr_t$ and $P(\tau)$ are the discounted return and the probability of a specific trajectory $\tau$, respectively, which can be computed as
\begin{equation}
P(\tau)=p(s_0)\prod_{t=0}^T\pi(a_t\mid s_t)P(s_{t+1}\mid s_t,a_t).
\end{equation}
Therefore, the agent policy $\pi$ can be improved by updating the policy parameters using gradient ascent. However, since the expected discounted return changes after the policy update, the previous trajectories cannot be utilized, leading to low data efficiency. To overcome the limitation, the PPO algorithm utilizes the importance sampling to improve the data efficiency. Specifically, each agent maintains two neural networks, i.e., the actor network and the critic network, which are used as the policy and to compute the expected discounted return, respectively. 
The actor network parameters are updated by the loss function 
\begin{equation}
\begin{aligned}
&\mathcal{L}_\text{actor}(\boldsymbol{\theta}_\text{a})=-\sum_{bs}\min\left(\frac{\pi(a_t|s_t;\boldsymbol{\theta}_\text{a})}{\pi(a_t|s_t;\boldsymbol{\theta}_\text{a}^\text{old})}A(s_t;\boldsymbol{\theta}_\text{v}),\right.\\ 
&\left.\text{clip}\Big(\frac{\pi(a_t|s_t;\boldsymbol{\theta}_\text{a})}{\pi(a_t|s_t;\boldsymbol{\theta}_\text{a}^\text{old})},1-\epsilon,1+\epsilon\Big)A(s_t;\boldsymbol{\theta}_\text{v})\right),
\label{PPOloss}
\end{aligned}
\end{equation}
where $bs$ is the batch size and $\text{clip}(\cdot, \cdot, \cdot)$ is the clipping function that clips the first input into the range determined by the second and third inputs, i.e., $[1-\epsilon, 1+\epsilon]$, with $\epsilon$ as the clipping factor, usually set to 0.2. Moreover, $\boldsymbol{\theta}_\text{a}^\text{old}$ and $\boldsymbol{\theta}_\text{a}$ are the old and new policy parameters, respectively. Therefore, the new policy parameters $\boldsymbol{\theta}_\text{a}$ can be updated several times using the trajectories generated by the interaction of the old policy $\boldsymbol{\theta}_\text{a}^\text{old}$ with the environment to improve the data efficiency. Furthermore, $A(s_t;\boldsymbol{\theta}_\text{v})$ denotes the advantages of each action relative to the old policy in a specific state $s_t$, given by
\begin{equation}
\label{advantage}
A(s_t;\boldsymbol{\theta}_\text{v})=\sum\limits_{l=0}^T(\gamma\lambda)^l(r_t+V(s_t;\boldsymbol{\theta}_\text{v})-V(s_{t+1};\boldsymbol{\theta}_\text{v})),
\end{equation}
where $\lambda$ is a parameter to trade off between the bias and variance of the advantage function, and $V(s_t;\boldsymbol{\theta}_\text{v})$ is the critic network value, parameterized by $\boldsymbol{\theta}_\text{v}$ and can be update by the loss function 
\begin{equation}
\label{critic}
\mathcal{L}_\text{critic}(\boldsymbol{\theta}_\text{v})=\sum_{bs}\big(r_t+\gamma V(s_{t+1};\boldsymbol{\theta}_\text{v})-V(s_t;\boldsymbol{\theta}_\text{v})\big)^2.
\end{equation}

Building upon the distributed hierarchical decision scheme and the PPO algorithm described above, we introduce the HMARL algorithm to train the low- and high-level policy. Specifically, each agent maintains two replay buffers to train the low- and high-level policies, respectively. As illustrated in Fig.~\ref{decision phase}, after each agent selects an option $o^i_t$ and performs an action $a^i_t$ at time step $t$, the combination of these joint options and actions interacting with the environment yields a reward $r_t$. Subsequently, the low- and high-level policy replay buffers store the transitions 
\begin{equation}
\label{trans_low}
(\tau^i_t,\boldsymbol{m}_t,o^i_t,a^i_t,r_{t,\text{tot}},r^i_{ind},\pi_i(a^i_t|o^i_t,\tau^i_t,\boldsymbol{m}_t;\boldsymbol{\theta}_{\pi,\text{old}}^i)),
\end{equation}
and
\begin{equation}
\label{trans_high}
(\tau^i_t,\boldsymbol{m}_t,o^i_t,\sum\limits_{t}^{t+E_0-1}r_{t,\text{tot}}, \sum\limits_{t}^{t+E_0-1}r_{t,\text{ind}}^i, \mu_i(o^i_t|\tau^i_t,\boldsymbol{m}_t;\boldsymbol{\theta}_{\mu,\text{old}}^i)),
\end{equation}
respectively, where $E_0$ is the number of packets each AP can transmit during single TXOP, and $\boldsymbol{\theta}_{\pi,\text{old}}^i$ and $\boldsymbol{\theta}_{\mu,\text{old}}^i$ are the old low- and high-level policy parameters, respectively. 
Moreover, the parameters of the encoder, low- and high-level policy of all agents, $\boldsymbol{\theta}_\text{m}^i$, $\boldsymbol{\theta}_\pi^i$, and $\boldsymbol{\theta}_\mu^i, \forall i \in \mathcal{K}$ are collectively denoted as $\boldsymbol{\theta}_\text{m}$, $\boldsymbol{\theta}_\pi$, and $\boldsymbol{\theta}_\mu$, respectively. As a result, agents can update the low- and high-level policy with the encoder network by the loss function
\begin{equation}
\label{actor_loss}
\mathcal{L}_\text{actor}(\boldsymbol{\theta}_\mu,\boldsymbol{\theta}_\pi, \boldsymbol{\theta}_\text{m})=\mathcal{L}_\text{low}(\boldsymbol{\theta}_\pi, \boldsymbol{\theta}_\text{m})+\mathcal{L}_\text{high}(\boldsymbol{\theta}_\mu, \boldsymbol{\theta}_\text{m}),
\end{equation}
where $\mathcal{L}_\text{low}(\boldsymbol{\theta}_\pi, \boldsymbol{\theta}_\text{m})$ and $\mathcal{L}_\text{high}(\boldsymbol{\theta}_\mu, \boldsymbol{\theta}_\text{m})$ are utilized to update the low- and high-level policy with the encoder network for all agents, respectively. To simplify the notation, we replace $\pi_i(a^i_t|o^i_t,\tau^i_t,\boldsymbol{m}_t(\boldsymbol{\theta}_m);\boldsymbol{\theta}^i_{\pi})$, $\pi_i(a^i_t|o^i_t,\tau^i_t,\boldsymbol{m}_t(\boldsymbol{\theta}_{m,\text{old}}); \boldsymbol{\theta}^i_{\pi,\text{old}})$, $\mu_i(o^i_t|\tau^i_t,\boldsymbol{m}_t(\boldsymbol{\theta}_m); \boldsymbol{\theta}^i_\mu)$ and $\mu_i(o^i_t|\tau^i_t,\boldsymbol{m}_t(\boldsymbol{\theta}_{m,\text{old}}); \boldsymbol{\theta}^i_{\mu,\text{old}})$ with $\pi_i(\boldsymbol{\theta}^i_\pi)$, $\pi_i(\boldsymbol{\theta}^i_{\pi,\text{old}})$, $\mu_i(\boldsymbol{\theta}^i_\mu)$ and $\mu_i(\boldsymbol{\theta}^i_{\mu,\text{old}})$, respectively.
Therefore, $\mathcal{L}_\text{low}(\boldsymbol{\theta}_\pi, \boldsymbol{\theta}_\text{m})$ and $\mathcal{L}_\text{high}(\boldsymbol{\theta}_\mu, \boldsymbol{\theta}_\text{m})$ are defined as 

\begin{equation}\scalebox{0.8}{$
\begin{aligned}
\mathcal{L}_\text{low}(\boldsymbol{\theta}_\pi, \boldsymbol{\theta}_\text{m})=&-\sum_{bs}\sum_{i}\min\Big(\frac{\pi_i( \boldsymbol{\theta}^i_\pi)}{\pi_i(\boldsymbol{\theta}^i_{\pi,\text{old}})}A(o^i_t,\tau^i_t,\boldsymbol{m}_t(\boldsymbol{\theta}_\text{m}); \boldsymbol{\theta}^i_{\pi, \text{V}}),\\
&\text{clip}\big(\frac{\pi_i(\boldsymbol{\theta}^i_\pi)}{\pi_i(\boldsymbol{\theta}^i_{\pi,\text{old}})},1-\epsilon,1+\epsilon\big)A(o^i_t,\tau^i_t,\boldsymbol{m}_t(\boldsymbol{\theta}_\text{m}); \boldsymbol{\theta}^i_{\pi, \text{V}})\Big),
\end{aligned}$}
\label{lowloss}
\end{equation}
and 
\begin{equation}\scalebox{0.8}{$
\begin{aligned}
\mathcal{L}_\text{high}(\boldsymbol{\theta}_\mu, \boldsymbol{\theta}_\text{m})=&-\sum_{bs}\sum_{i}\min\Big(\frac{\mu_i(\boldsymbol{\theta}^i_\mu)}{\mu_i(\boldsymbol{\theta}^i_{\mu,\text{old}})}A(\tau^i_t,\boldsymbol{m}_t(\boldsymbol{\theta}_\text{m}); \boldsymbol{\theta}^i_{\mu, \text{V}}),\\
&\text{clip}\big(\frac{\mu_i(\boldsymbol{\theta}^i_\mu)}{\mu_i(\boldsymbol{\theta}^i_{\mu,\text{old}})},1-\epsilon,1+\epsilon\big)A(\tau^i_t,\boldsymbol{m}_t(\boldsymbol{\theta}_\text{m}); \boldsymbol{\theta}^i_{\mu, \text{V}})\Big),
\end{aligned}$}
\label{highloss}
\end{equation}
where $A(o^i_t,\tau^i_t,\boldsymbol{m}_t(\boldsymbol{\theta}_m); \boldsymbol{\theta}^i_{\pi,\text{V}})$ and $A(\tau^i_t,\boldsymbol{m}_t(\boldsymbol{\theta}_\text{m}); \boldsymbol{\theta}^i_{\mu, \text{V}})$ are the advantage functions for the low- and high-level policy, parameterized by $\boldsymbol{\theta}^i_{\pi,\text{V}}$ and $\boldsymbol{\theta}^i_{\mu, \text{V}}$, respectively. As mentioned above, our goal is to maximize network throughput while maintaining fairness among agents. However, these two goals may be contradictory, as maximizing network throughput may cause a decrease in fairness and vice versa, which indicates multi-objective problems require trade-off. Therefore, instead of summing the total reward $r_{t,\text{tot}}$ and individual reward $r^i_{t,\text{ind}}$ to compute the advantage for each agent, which could result in suboptimal solutions, we adopt the multi-critic single policy (MCSP) \cite{han2024multi, nguyen2021prioritized}. The MCSP method employs multiple critics to evaluate an agent, which allows for a more nuanced assessment of both total throughput and fairness contributions. Specifically, we set the advantage function of the low-level policy for agent $i$ as 
\begin{equation}\
\begin{aligned}
A(o^i_t,\tau^i_t,\boldsymbol{m}_t(\boldsymbol{\theta}_m); \boldsymbol{\theta}^i_{\pi, \text{V}})=&\omega_1A^i_\text{tot}(o^i_t,\tau^i_t,\boldsymbol{m}_t(\boldsymbol{\theta}_m); \boldsymbol{\theta}^i_{\pi, \text{V}, \text{tot}})+\\
&\omega_2A^i_\text{ind}(o^i_t,\tau^i_t,\boldsymbol{m}_t(\boldsymbol{\theta}_m); \boldsymbol{\theta}^i_{\pi, \text{V},\text{ind}}),
\end{aligned}
\end{equation}
where $\omega_1$ and $\omega_2$ are positive weights to balance the two objectives. $A^i_\text{tot}(o^i_t,\tau^i_t,\boldsymbol{m}_t(\boldsymbol{\theta}_m); \boldsymbol{\theta}^i_{\pi, \text{V}, \text{tot}})$ and $A^i_\text{ind}(o^i_t,\tau^i_t,\boldsymbol{m}_t(\boldsymbol{\theta}_m); \boldsymbol{\theta}^i_{\pi, \text{V}, \text{ind}})$ are the advantages for the two objectives obtained with two critic networks with parameters $\boldsymbol{\theta}^i_{\pi, \text{V}, \text{tot}}$ and $\boldsymbol{\theta}^i_{\pi, \text{V}, \text{ind}}$, calculated by (\ref{advantage}), respectively. The parameters $\boldsymbol{\theta}^i_{\pi, \text{V}, \text{tot}}$ and $\boldsymbol{\theta}^i_{\pi,\text{V}, \text{ind}}$ together form the parameters $\boldsymbol{\theta}^i_{\pi,\text{V}}$. We can update the parameters $\boldsymbol{\theta}^i_{\pi, \text{V}}$ with a loss function of the same form as loss (\ref{critic}). Moreover, the advantage function of the high-level policy $ A(\tau^i_t,\boldsymbol{m}_t(\boldsymbol{\theta}_\text{m}); \boldsymbol{\theta}^i_{\mu, \text{V}})$ and its parameters $\boldsymbol{\theta}^i_{\mu, \text{V}}$ are similarly defined and updated. 

Overall, the training procedure of the proposed HMARL algorithm is summarized in Algorithm~\ref{HMARLalgorithm}.
\begin{algorithm}[!t]
\caption{HMARL Algorithm} 
\label{HMARLalgorithm}
\SetKwData{Left}{left}\SetKwData{This}{this}\SetKwData{Up}{up}
 \SetKwFunction{Union}{Union}\SetKwFunction{FindCompress}{FindCompress}
 \SetKwInOut{Input}{Initialization}\SetKwInOut{Output}{output}
\Input{$t =0, \epsilon, E=E_0$, $\boldsymbol{\theta}_\text{m}$, $\boldsymbol{\theta}_\pi,\boldsymbol{\theta}_\mu$, $\boldsymbol{\theta}^i_{\pi,\text{V}}$, $\boldsymbol{\theta}^i_{\mu,\text{V}}$, $\tau^i_0=\tau_0, m^i_0=m_0, \forall i \in \mathcal{K}$,\\ phase = contention.}

\While{$t<T$}{
    \If{\textup{phase = contention phase}}{
        Each AP adopts the CSMA/CA mechanism to contend for a TXOP.\\
        \If{\textup{One AP occupies the TXOP}}{
            phase = polling, $E = E_0$.
        }
    }
    \If{\textup{phase = polling}}{
    \For{\textup{Agent} $i=1,2...,K$}{
        \eIf{\textup{Agent $i$ is sharing AP}}{
            \eIf{\textup{Polling is not over}}{
                AP $i$ polls other APs whether or not to participate in the CSR.\\
            }{
                AP $i$ obtains $\tau^i_t$ and sends the collective information $\{m^i_t, \forall i \in \mathcal{K}\}$ to shared APs.
            }
        }{
            AP $i$ obtains $\tau^i_t$ and sends information $m^i_t$ to the sharing AP.\\
            phase = decision.
        }
    }}
    \If{\textup{phase = decision}}
    {
    \For{\textup{Agent} $i=1,2...,K$}{
    \eIf{\textup{Agent $i$ has not taken a decision}}{
            \If{\textup{$E = E_0$}}{
                Agent $i$ chooses an option $o^i_t$ and an action $a^i_t$ to transmit.
            }
            \eIf{\textup{$E = 0$}}{
                Store the transition (\ref{trans_high}) to the high-level policy replay buffer.\\
                phase = contention.
            }{
                Agent $i$ chooses an action $a^i_t$ to transmit.
            }
        }{  
            \eIf{\textup{Agent $i$ dose not finish transmission}}{
                Agent $i$ transmits.
            }{  
                $E \leftarrow E - 1$. Update the reward $r_t$, and store the transition (\ref{trans_low}) to the low-level policy replay buffer.
            }
        }
    }
    }

\If{$t$ \textup{mod} $N_c$ = 0}{
    Each agent samples experiences from the low- and high-level replay buffers and compute (\ref{actor_loss}).
    Update $\boldsymbol{\theta}_\text{m}$, $\boldsymbol{\theta}_\pi,\boldsymbol{\theta}_\mu$, $\boldsymbol{\theta^i}_{\pi,\text{V}}$, $\boldsymbol{\theta}^i_{\mu,\text{V}}$ by gradient descent. 
}
$t\leftarrow t+1$
}
\end{algorithm}

\section{Performance Evaluation}
This section provides simulation results of the proposed HMARL algorithm in different OBSS representative topologies, where each BSS consists of an AP and several STAs. In addition, we conducted an ablation study to evaluate the effectiveness of broadcast information and hierarchical architecture, together with a robustness assessment to assess the stability of the HMARL algorithm. Furthermore, we analyze the improvement of the designed reward function on network fairness in a particular topology. Additionally, our investigation extends to the coexistence scenarios of the learning-based APs and the legacy APs, verifying that our algorithm does not degrade the performance of existing legacy systems.

In the following, we first introduce the simulation setup and after that the detailed experimental results will be presented. 

\subsection{Simulation Setup}
\subsubsection{Simulation Scenario}
\begin{figure}[!t]
\centering
\includegraphics[width=0.49\textwidth]{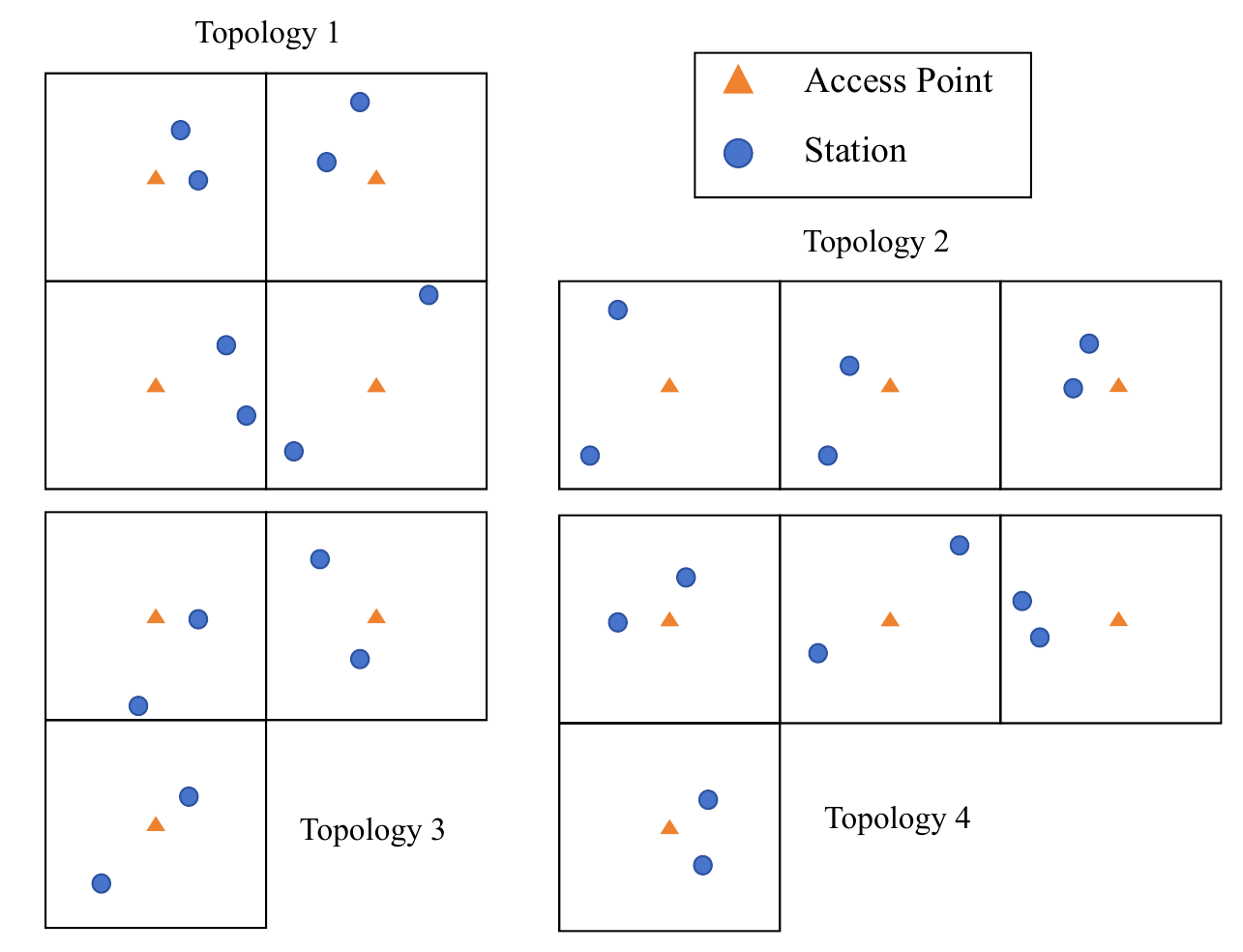}
\caption{Representative topology information: The orange triangle represents the AP and the blue circles represent the STAs}
\label{topology}
\end{figure}

We consider four representative topologies in a residential scenario for the simulation. As illustrated in Fig.~\ref{topology}, each BSS consists of one AP and two STAs, where the orange triangle represents the AP and the blue circles represent the STAs. Each AP is located in the center of a room and the positions of the STAs are initialized randomly. Each room measures 10 meters in both length and width, with walls separating the rooms. The detailed simulation parameters of the OBSS scenarios are shown in Table~\ref{Simulation Parameters}. Note that the choice of -100 dBm actually means zero transmit power.
We use the following metrics to evaluate the performance of the proposed HMARL algorithm.
\begin{itemize}
\item Throughput: The ratio of successful transmission slots to total time slots.
\item Mean Delay: The average delay of all successfully transmitted packets, measured as the number of time slots from packet generation to successful transmission.
\item Delay Jitter: The variance in the delay of all successfully transmitted packets.
\end{itemize}

\subsubsection{Benchmarks}
We compare the proposed HMARL algorithm against the CSMA/CA mechanism, where CCA thresholds are set to -82 dBm.
Moreover, we benchmark the proposed HMARL algorithm against two MARL-based algorithms, demonstrating that our approach provides high throughput compared to adjust CCA thresholds and transmit power.
\begin{itemize}
\item MARLCCA: An algorithm based on distributed MARL \cite{yan2024multi}. Each AP broadcasts information every 0.5 seconds and then utilizes the received information and its own observations as inputs to a policy network to adjust the CCA thresholds in the range of $[-82,-62]$ dBm to optimize spatial reuse. In addition, according to the BSS coloring mechanism \cite{wifi6standard}, the transmit power $P$ of AP is set as $P=-82-\textup{CCA}$ dBm.
\item HRLCCA+Power: A hierarchical MARL algorithm based on the MARLCCA algorithm to jointly optimize the CCA threshold and the transmit power \cite{huang2023hierarchical}. The high-level policy adjusts the CCA thresholds and the low-level policy adjusts the transmit power.
\end{itemize}
\begin{table}[!t]
\caption{Simulation Parameters}
\label{Simulation Parameters}
\centering
    \begin{tabular}{|c|c|}
        \hline
        \textbf{Parameters} & \textbf{value} \\
        \hline
        Transmit power  & [20, 15, 10, 5, -100] dBm \\
        \hline
        Shadowing standard deviation & 3 dB\\
        \hline
        Topology room length and width & 10 m, 10 m \\
        \hline
        Break distance & 5 m \\
        \hline
        MCS & 5 \\
        \hline
        Time slot & 9 $\upmu$s \\
        \hline
        Packet length & 1080 $\upmu$s \\
        \hline
        ACK length & 36 $\upmu$s \\
        \hline
        Number of transmissions per TXOP $E_0$ & 3 \\
        \hline
        CSMA/CA contention window range & [31, 1023] \\
        \hline
        Default CCA threshold & -82 dBm \\
        \hline
        Traffic type & Saturated Poisson traffic \\
        \hline
        Large-scale fading model & Residential scenario in \cite{residential}\\
        \hline
        Small-scale fading model & Nakagami-$m$ \\
        \hline
        $m$  & 1.5\\
        \hline
        Small-scale fading update & Every packet \\
        \hline
    \end{tabular}
\end{table}

\begin{table}[!t]
\caption{Hyperparameters}
\label{Hyper-Parameters}
\centering
    \begin{tabular}{|c|c|}
        \hline
        \textbf{Parameters} & \textbf{value} \\
        \hline
        Discount factor $\gamma$ & 0.5 \\
        \hline
        $\lambda$ & 0.95 \\
        \hline
        Agent update interval & 20 TXOPs \\
        \hline 
        Learning rate & $1 \times 10^{-4}$ \\
        \hline
        Observe history length &5 \\
        \hline
        Dimension of information & 4 \\
        \hline
        Activation function & ReLU \\
        \hline
        Neurons of PPO agent critic & 250, 120, 120\\
        \hline
        Neurons of PPO agent actor & 250, 120, 120\\
        \hline
        Optimizer & RMSProp \\
        \hline
    \end{tabular}
\end{table}
\subsubsection{Hyperparameters of HMARL}
The hyperparameters of the MAHRL algorithm are shown in Table~\ref{Hyper-Parameters}. The actor and critic networks are both composed of three multilayer perceptron layers with the ReLU as the activation function for each layer. The RMSProp optimizer is utilized to update the network parameters every 20 TXOP intervals. Furthermore, the information extracted by the encoder consists of 4 float values with each float being 32 bits. The total amount of information per AP in one  second under saturated traffic is approximately $\frac{1}{3\times(1080+36)\times10^{-6}}\times32\times4\approx 0.038$ Mbps, which is significantly smaller than the data rate $68.8$ Mbps of the corresponding MCS.

\subsection{Simulation Results of Representative Topologies}


\begin{figure}[!t]
    \centering
    \begin{minipage}[t]{0.48\textwidth}
        \centering
        \subfigure[Throughput (\%)]{
            \includegraphics[width=3.3in]{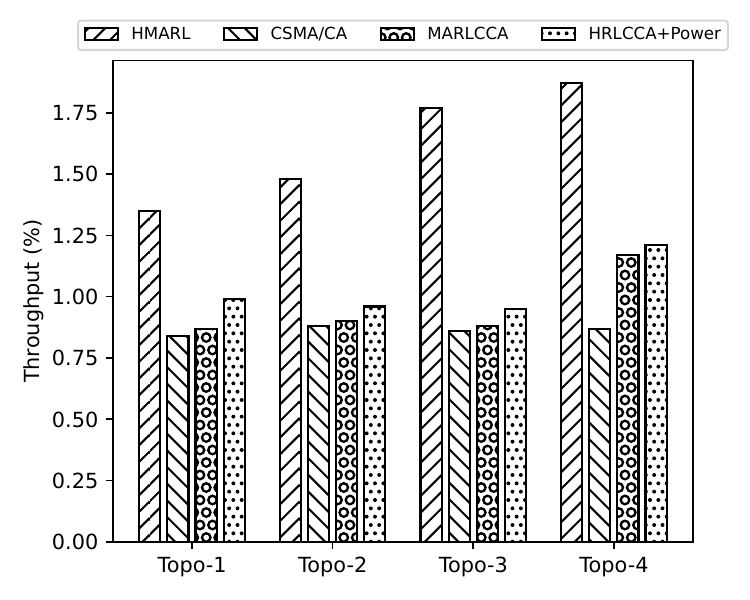}
            \label{throughput}
        }
    \end{minipage}

    \begin{minipage}[t]{0.48\textwidth}
        \centering
        \subfigure[Mean delay (s)]{
            \includegraphics[width=3.3in]{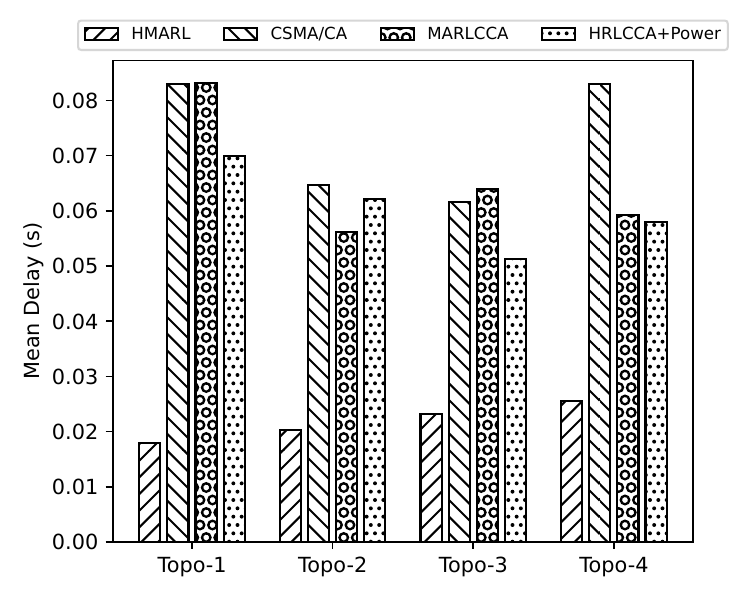}
            \label{mean delay}
        }
    \end{minipage}
    
    \begin{minipage}[t]{0.48\textwidth}
        \centering
        \subfigure[Delay jitter ($\text{s}^2$)]{
            \includegraphics[width=3.3in]{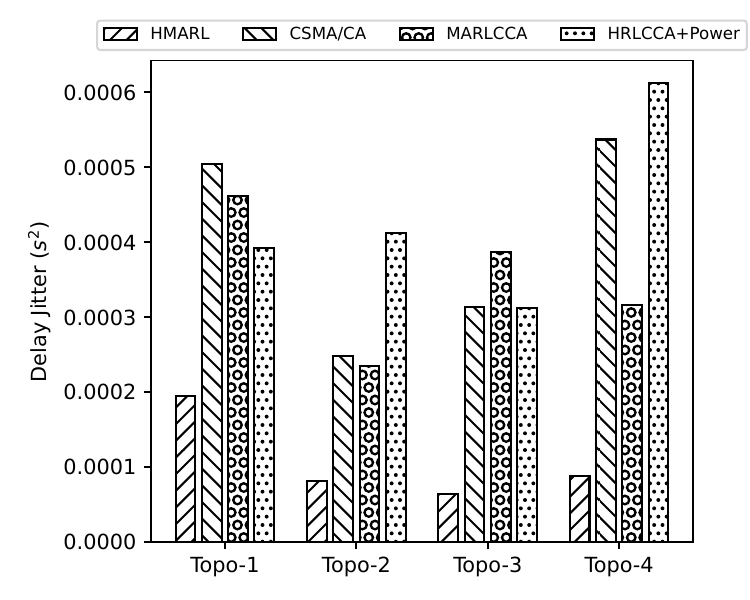}
            \label{delay jitter}
        }
    \end{minipage}

    \caption{Performance comparison under saturated Poisson traffic.}
    \label{representative topo}
\end{figure}
Fig.~\ref{representative topo} illustrates the throughput, mean delay, and delay jitter performance in representative topologies for different algorithms. 

\textbf{Throughput Performance:} From Fig.~\ref{representative topo}~\subref{throughput}, the throughput performance of HMARL, MARLCCA, and HRLCCA+Power algorithms all outperforms the CSMA/CA mechanism, demonstrating the effectiveness of these RL-based approaches in improving network efficiency. Among them, HRLCCA+Power achieves better performance than MARLCCA, benefiting from its hierarchical structure that enables joint optimization of CCA thresholds and transmit power. However, our proposed HMARL algorithm achieves the highest throughput compared to other RL-based algorithms. Unlike CCA-based methods, which rely on passive adjustments that may not fully exploit spatial reuse opportunities, our approach utilizes multi-AP collaboration, where each AP has the opportunity to participate in spatial reuse at each TXOP, and the interference is minimized as much as possible through STA selection and power allocation, thus obtaining the highest throughput. 
Specifically, the proposed HMARL algorithm increases the throughput performance by approximately $0.50, 0.60, 0.91$, and $0.95$ compared to CSMA/CA across the four representative topologies, respectively. 

\textbf{Mean Delay:} From Fig.~\ref{representative topo}~\subref{mean delay}, the mean delay of the HMARL algorithm is only one-half to one-third of the mean delay achieved by the CSMA/CA mechanism, which outperforms other RL-based algorithms. This reduction in mean delay can be attributed to the HMARL algorithm utilizing multi-AP collaboration, which allows packets to be transmitted in parallel most of the time. In contrast, the CSMA/CA mechanism primarily transmits packets sequentially, resulting in higher latency. Consequently, the HMARL algorithm achieves a significant decrease in transmission delays. 

\textbf{Delay Jitter:}
From Fig.~\ref{representative topo}~\subref{delay jitter}, HMARL exhibits lower delay jitter relative to other algorithms. Reduced jitter indicates more consistent and predictable transmission performance, which is particularly beneficial for latency-sensitive applications.



\subsection{Ablation Study}
\begin{figure}[!t]
    \centering

    \begin{minipage}[t]{0.48\textwidth}
        \centering
        \subfigure[Throughput (\%)]{
            \includegraphics[width=3.2in]{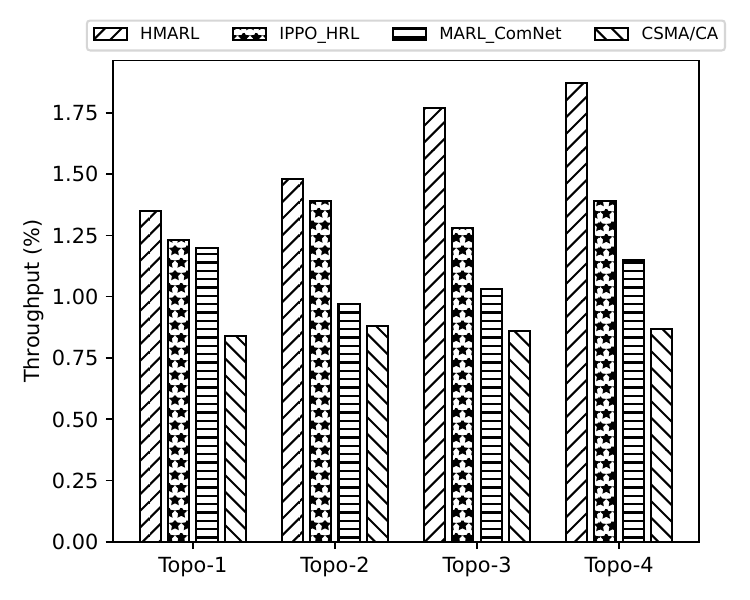}
            \label{ab throughput}
        }
    \end{minipage}

    \begin{minipage}[t]{0.48\textwidth}
        \centering
        \subfigure[Mean delay (s)]{
            \includegraphics[width=3.2in]{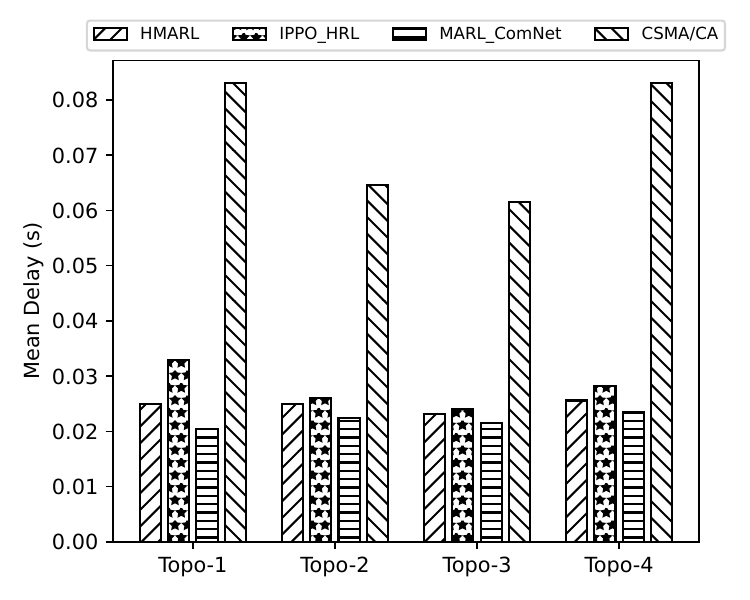}
            \label{ab mean delay}
        }
    \end{minipage}
    
    \begin{minipage}[t]{0.48\textwidth}
        \centering
        \subfigure[Delay jitter ($\text{s}^2$)]{
            \includegraphics[width=3.2in]{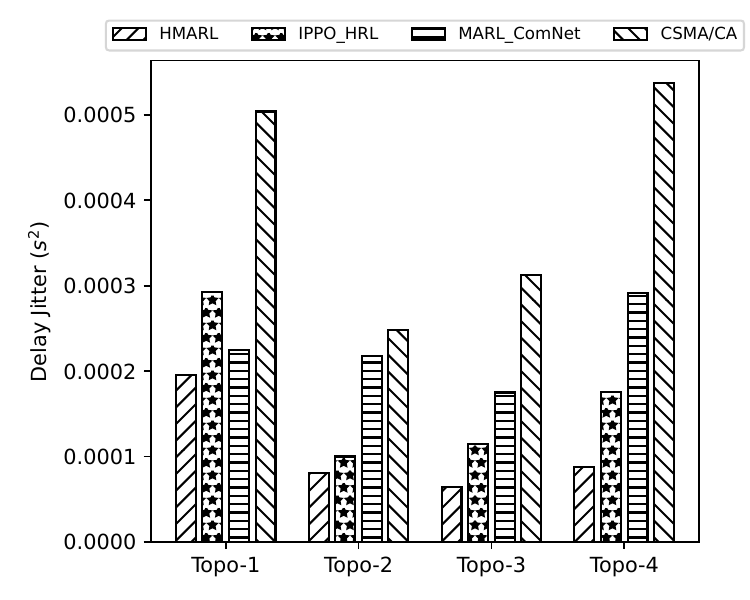}
            \label{ab delay jitter}
        }
    \end{minipage}

    \caption{Performance comparison between the HMARL algorithm and ablation schemes.}
    \label{ablation}
\end{figure}

In this subsection, we evaluate the effectiveness of the exchanged information among agents as well as the hierarchical architecture design for our proposed algorithm through two ablation experiments, namely IPPO\_HRL and MARL\_ComNet. Both ablation schemes share the same reinforcement learning components as the proposed HMARL algorithm and are evaluated under the representative topologies in Fig.~\ref{topology}. The two schemes are detailed below:
\begin{itemize}
\item {\bfseries IPPO\_HRL:} This variant removes the information exchange mechanism in the polling phase of our proposed CSR framework. As a result, the agents can only utilize their local observations to make decisions based on the hierarchical architecture for STA selection and power control. 
\item {\bfseries MARL\_ComNet:} In contrast to IPPO\_HRL, the MARL\_ComNet algorithm retains the information exchange mechanism in the polling phase, but removes the hierarchical architecture, and instead employs only one policy network for STA selection and power allocation in the decision phase. Thus, each agent is equipped with a single flat policy network and a communication encoder, jointly used for both STA selection and power control.
\end{itemize}

Fig.~\ref{ablation} compares the throughput, mean delay, and delay jitter performance comparison among HMARL, IPPO\_HRL, and MARL\_ComNet. Both IPPO\_HRL and MARL\_ComNet exhibit significant improvement across all three metrics compared with the CSMA/CA mechanism, demonstrating that both the two module designs can promote the cooperative spatial reuse of multiple APs. Moreover, Fig.~\ref{ablation}~\subref{ab throughput} shows that the HMARL algorithm further improves the throughput compared to the two ablation schemes, highlighting that the joint design of information exchange and hierarchical structure leads to enhanced coordination among APs and superior overall performance.

\subsection{Robustness and Coexistence Test}
To evaluate the robustness of the HMARL algorithm, we conduct a cross-topology experiment, where the model is trained in one topology and tested in another. Additionally, we evaluate the network performance when our proposed HMARL algorithm and the conventional CSMA/CA coexist in a network. Specifically, we train four learning-based APs with the HMARL algorithm in the scenario of topology 1 in Fig.~\ref{representative topo}, and then test the network performance under different numbers of learning-based APs and legacy APs coexistence in the same topology. 
Since legacy APs do not implement neural networks, they cannot exchange information with other learning-based APs or participate in the CSR. To address this, the missing information from the legacy APs for the learning-based AP is replaced with a zero vector of the same length as the input data of the neural network. For fairness, the learning-based APs and the legacy APs take up the same length of the TXOP.

Table~\ref{cross-experiment I} and \ref{cross-experiment II} show the simulation results of the two cross-topology experiments for the HMARL algorithm trained in topology 3 and tested in topology 2, and trained in topology 4 and tested in topology 1, respectively, where $(x/y)$ denotes the algorithm trained in topology $x$ and tested in topology $y$. From these two tables, despite performance degradation due to topology mismatch, HMARL still has substantial throughput and latency gains compared to the traditional CSMA/CA method, and the performance is comparable to the other two ablation schemes. These results suggest that the robustness of the HMARL algorithm in different topologies should be taken with a grain of salt: Within a reasonable region of topology change, the trained model is good, however, needs to be retrained if the test topology changes significantly, as the expected performance will eventually drop to an unacceptable level.

Table~\ref{coexistence} presents the simulation results of the coexistence of learning-based APs and legacy APs. In addition to the three metrics used before, we also introduce the average throughput of the legacy APs $\overline{\text{THP}_l}$ as a metric to evaluate whether the learning-based APs affect the performance of the legacy APs. Moreover, since legacy APs can not participate in CSR, the scenario with one learning-based AP is equivalent to the all-legacy case and is thus omitted.
From this table, as the number of learning-based APs increases, the network throughput increases and the mean delay decreases, due to that the more APs are able to participate in CSR. Furthermore, the average throughput $\overline{\text{THP}_l}$ of the legacy APs does not decrease when coexisting with learning-based APs compared to that of the scenario with all legacy APs, demonstrating that the HMARL algorithm is compatible with the conventional CSMA/CA mechanism. This compatibility arises from that the HMARL algorithm is based on the CSMA/CA mechanism for a compete-then-collaborate approach, and therefore does not conflict with traditional CSMA/CA mechanism.


\begin{table}[!t]
\caption{Simulation results of cross-experiment I}
\label{cross-experiment I}
\centering
\scriptsize 
    \setlength{\tabcolsep}{2pt} 
    \renewcommand{\arraystretch}{1.5} 
    \begin{tabular}{|c|c|c|c|}
        \hline
        \textbf{Algorithm} & \textbf{Throughput ($\%$)} &\textbf{Mean delay (s)} &\textbf{Delay jitter ($\text{s}^2$) } \\
        \hline
        HMARL (3/2) & $1.31$ & $0.0232$ & $8.92 \times 10^{-5}$ \\
        \hline
        HMARL (2/2) & $1.48$ & $0.0202$ & $8.12 \times 10^{-5}$ \\
        \hline
        IPPO\_HRL (2/2) & $1.39$ & $0.0261$ & $1.03 \times 10 ^{-4}$ \\
        \hline
        MARL\_ComNet (2/2) & $0.97$ & $0.0224$ & $2.18 \times10^{-4}$ \\
        \hline
        CSMA/CA & $0.84$ & $0.0646$ & $2.48 \times 10^{-4}$ \\
        \hline
    \end{tabular} 
\end{table}

\begin{table}[!t]
\caption{Simulation results of cross-experiment II}
\label{cross-experiment II}
\centering
\scriptsize 
    \setlength{\tabcolsep}{2pt} 
    \renewcommand{\arraystretch}{1.5} 
    \begin{tabular}{|c|c|c|c|}
        \hline
        \textbf{Algorithm} & \textbf{Throughput ($\%$)} &\textbf{Mean delay (s)} &\textbf{Delay jitter ($\text{s}^2$) } \\
        \hline
        HMARL (4/1) & $1.21$ & $0.0198$ & $1.33 \times 10^{-5}$ \\
        \hline
        HMARL (1/1) & $1.35$ & $0.0179$ & $1.03 \times 10^{-4}$ \\
        \hline
        IPPO\_HRL (1/1) & $1.23$ & $0.0331$ & $2.93 \times 10 ^{-4}$ \\
        \hline
        MARL\_ComNet (1/1) & $1.20$ & $0.0278$ & $2.25 \times10^{-4}$ \\
        \hline
        CSMA/CA & $0.87$ & $0.0830$ & $5.04 \times 10^{-4}$ \\
        \hline
    \end{tabular} 
\end{table}


\begin{table}[!t]
\caption{Simulation results of the coexistence of Learning-based APs and legacy APs.}
\label{coexistence}
\centering
\scriptsize 
\setlength{\tabcolsep}{2pt} 
\renewcommand{\arraystretch}{1.5} 
\begin{tabular}{|c|c|c|c|c|}
    \hline
    \textbf{\makecell{Learning-based\\STA number}} & $\overline{\text{THP}_l}$  & \textbf{Throughput (\%)} &\textbf{Mean delay (s)} &\textbf{Delay jitter ($\text{s}^2$)} \\
    \hline
    0 & $0.219$ & $0.874$ & $0.0732$ & $5.60 \times 10^{-4}$ \\
    \hline
    2 & $0.227$ & $1.05$ & $0.0634$ & $8.02 \times 10^{-4}$ \\
    \hline
    3 & $0.221$ & $1.28$ & $0.0406$ & $4.55 \times 10^{-4}$ \\
    \hline
    4 & $\backslash$ & $1.43$ & $0.0294$ & $1.73 \times 10^{-4}$ \\
    \hline
\end{tabular} 
\end{table}

\subsection{Fairness Analysis}
\begin{figure}[!t]
\centering
\includegraphics[width=0.49\textwidth]{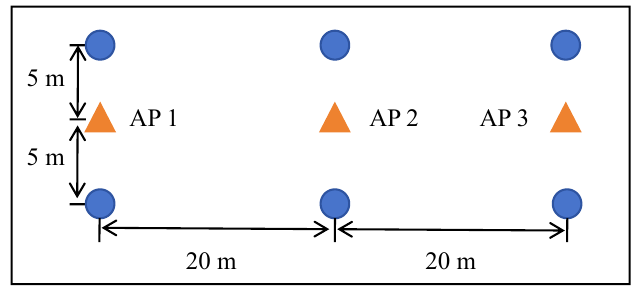}
\caption{A particular topology. The orange triangle represents the AP and the blue circles represent the STAs.}
\label{special topo}
\end{figure}

\begin{figure}[!t]
    \begin{minipage}[t]{0.9\textwidth}
        \subfigure[Real-time throughput using $r_{t,\text{tot}}$ and $r^i_{t,\text{ind}}$.]{
            \includegraphics[width=3.5in]{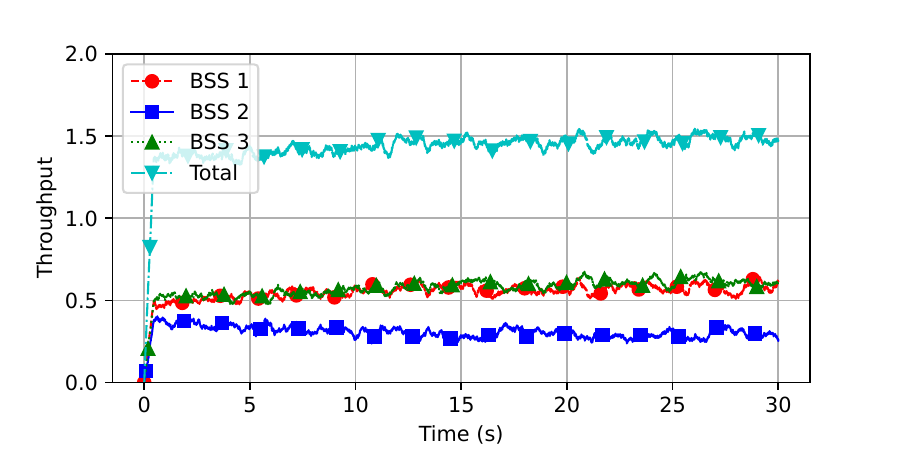}
            \label{reward1}
        }
    \end{minipage}%

    \begin{minipage}[t]{0.9\textwidth}
        \subfigure[Real-time throughput using $r_{t,\text{tot}}$.]{
            \includegraphics[width=3.5in]{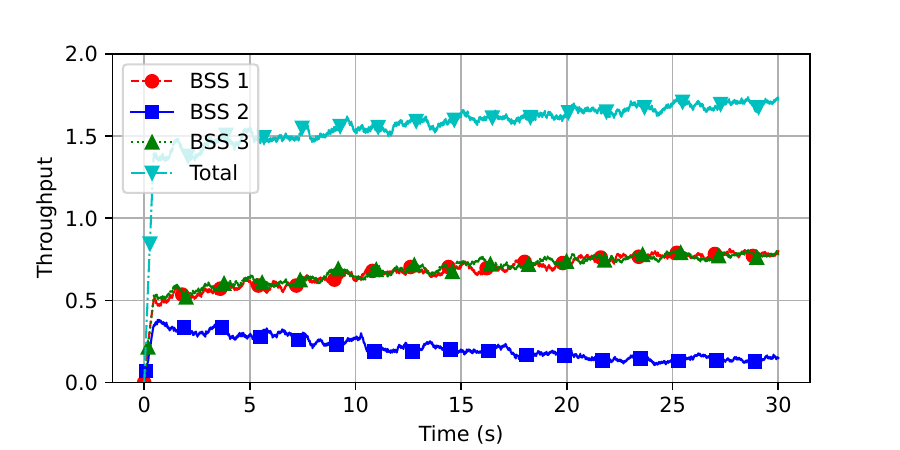}
            \label{reward2}
        }
    \end{minipage}%

    \begin{minipage}[t]{0.9\textwidth}
        \subfigure[Real-time throughput using $r_t=|\mathcal{S}_t|$.]{
            \includegraphics[width=3.5in]{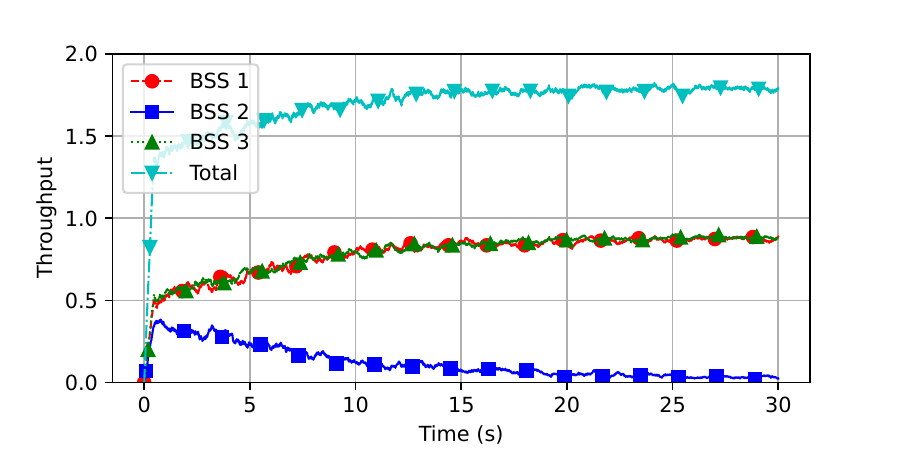}
            \label{reward3}
        }
    \end{minipage}%

    \caption{Performance comparison between the HMARL algorithm and ablation schemes.}
    \label{fairness}
\end{figure}
In this subsection, we analyze how the designed total reward $r_{t,\text{tot}}$ and individual reward $r^i_{t,\text{ind}}$ promote fairness in throughput across each BSS, i.e., the first constraint in the optimization problem (\ref{problem}). As illustrated in Fig.~\ref{special topo}, we consider a symmetric topology consisting of three APs for simulation, each associated with two STAs, where the distance between each STA and its corresponding AP is 5 m, and the distance between each AP is 20 m. We first train models using the designed reward functions $r_{t,\text{tot}}$ and $r^i_{t,\text{ind}}$, which are described in Section~\ref{Section III}. Then we utilize two reward functions $r_{t,\text{tot}}$ and $r_t=|\mathcal{S}_t|$, respectively, to train models in the same topology, where $\mathcal{S}_t$ represents the set of APs that transmit successfully at time step $t$. Subsequently, we can utilize the real-time throughput of each BSS during training using the three reward functions to analyze the fairness at convergence, where the real-time throughput is defined as the ratio of the successful transmission slots over the past 0.5 seconds.

Fig.~\ref{fairness} shows the real-time throughput during training using the three different reward functions. From Fig.~\ref{fairness}~\subref{reward1} to \subref{reward3}, the real-time throughput of AP 1 and AP 3 exhibit similar behavior across all reward functions, converging to similar performance. The total network throughput converges to approximately 1.5, 1.6, and 1.8, respectively, under the three reward settings.
However, the throughput of AP 2—located in a high-interference region—is best preserved when trained using the rewards $r_{t,\text{tot}}$ and $r^i_{t,\text{ind}}$ together, although it remains slightly lower than that of AP 1 and AP 3. In contrast, with $r_{t,\text{tot}}$ and $r_t=|\mathcal{S}_t|$, AP 2 is more severely starved of channel access.
This is because when using the reward function $r_t=|\mathcal{S}_t|$, the agents learn in the direction of maximizing the network throughput. Although fairness consideration is given for the reward function $r_{t,\text{tot}}$, the system eventually converges to maximize throughput, because even if the agent with the lower throughput does not transmit successfully, all agents can still obtain a positive reward value, as analyzed in Section~\ref{Section III}. However, when combining the rewards $r_{t,\text{tot}}$ and $r^i_{t,\text{ind}}$, it is possible to better restrict AP 1 and AP 3, which are in a weak interference position, from occupying too much of the channel, thereby giving AP 2, which is in a strong interference position, certain transmission opportunities, and ensuring certain fairness. In conjunction with the observations, we conclude that the transmission opportunities of the AP in a strong interference position will be sacrificed when the reward function is designed to maximize the network throughput. In contrast, when fairness among APs needs to be considered, a certain amount of overall throughput performance is sacrificed so that the lowest throughput AP is not too low. Therefore, maximizing throughput and considering fairness may be two conflicting objectives that need to be carefully traded off. 

\begin{figure}[!t]

        
    \begin{minipage}[t]{0.9\textwidth}
        \subfigure[Averaged reward]{
            \includegraphics[width=3.5in]{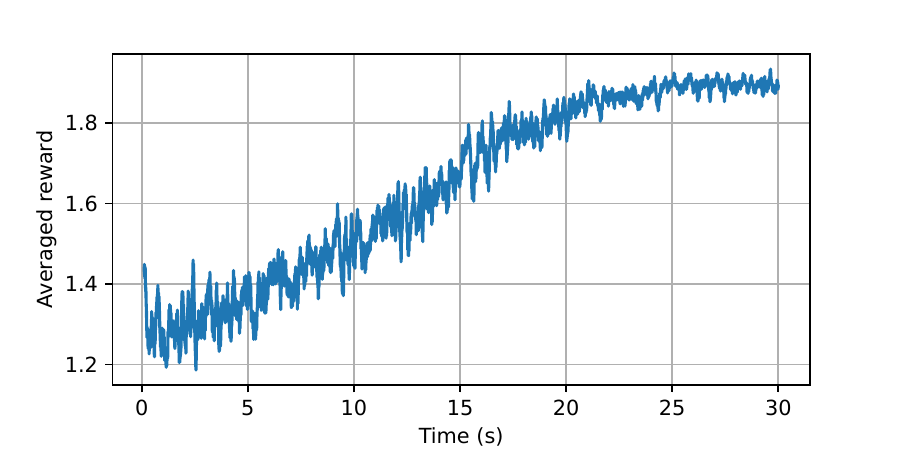}
            \label{averaged reward}
        }
    \end{minipage}

    \caption{Averaged reward.}
    \label{convergence}
\end{figure}
Fig.~\ref{convergence} illustrates the total reward with increasing training times in the particular topology shown in Fig.~\ref{special topo} to show the convergence behavior or the proposed HMARL algorithm, where the total reward is averaged over the past 100 time steps. From this figure, the total reward increases as the training continues and converges at approximately 20 seconds, demonstrating the effectiveness of the proposed HMARL algorithm. 

\section{Conclusion}
In this paper, we have introduced a hierarchical MARL-based algorithm to enhance CSR in the OBSS scenarios for next-generation WLANs, which adopts a fully distributed paradigm for training and testing. Additionally, 
simulation results demonstrate that the proposed HMARL algorithm enhances both network throughput and fairness across four representative topologies as well as a specific topology. Ablation studies further highlight the critical roles of broadcast information and the hierarchical architecture in achieving these gains. Furthermore, the algorithm is tested in scenarios where learning-based APs coexist with legacy APs, confirming that it does not degrade the performance of traditional CSMA/CA systems. Future work will explore the extension of the proposed HMARL algorithm to the uplink scenarios in OBSS scenarios.

\bibliography{references}
\bibliographystyle{IEEEtran}

\vfill

\end{document}